\documentstyle[sprocl,psfig]{article}
\def\Journal#1#2#3#4{{#1} {\bf #2}, #3 (#4)}

\def\be{\begin{equation}}
\def\ee{\end{equation}}
\def\bea{\begin{eqnarray}}
\def\eea{\end{eqnarray}}

\begin{document}
\title{Evolution of the morphological luminosity distributions within rich 
clusters ($0.0 < z < 0.55$)}
\author{ S.P.Driver, W.J.Couch }
\address{School of Physics, University of New South Wales, \\
Sydney, NSW 2052, AUSTRALIA}
\author{ S.C.Odewahn, R.A.Windhorst }
\address{Department of Physics and Astronomy, \\ 
Arizona State University, Tempe, AZ 85287-1504, USA}

\maketitle\abstracts{We demonstrate the ability to recover
morphological luminosity distributions (LDs) within medium redshift clusters
($z \sim 0.55$) based on {\it Hubble Space Telescope} WFPC2 observations.
We postulate that a detailed survey of the morphological LDs in local, low 
and medium redshift clusters may provide strong constraints on the modes
of galaxy evolution in rich clusters. Preliminary results suggest 
that in clusters, as also seen in the field, very strong evolution
({\it i.e.} $\Delta m \approx 2.5$ mags since $z = 0.55$) is 
occurring in the late-type spiral and irregular populations.}

\section{Introduction}
The Canada-France-Redshift-Survey (CFRS) of field galaxies has demonstrated 
that by recovering the optical luminosity distributions (LDs) of galaxies over 
a wide range of epochs it is possible to place strong constraints on the
amount of stellar evolution for various galaxy types~\cite{lilly}. Through 
these surveys, 
the results of morphological number count 
studies~\cite{dwg}~\cite{kgb}~\cite{dwokgr}
and the surface photometry of MgII absorption systems~\cite{steidel}, 
we now have a consistent picture of the passive 
evolution of elliptical and early-type galaxies and the rapid evolution of
late-type systems over the redshift range $0.0 < z < 1.0$.

Here we introduce an analogous program to trace the evolution of the galaxy
LDs within rich cluster environments. Recovering the LDs
within rich clusters is much simpler than for field galaxies as 
clusters represent the ideal prepackaged volume-limited sample. Essentially 
to recover a cluster's LF all that is required is the redshift
of the cluster and the subtraction of the mean field number-counts from the 
number-counts along the cluster sight-line. This technique has now
been applied to a number of clusters and more details on the subtleties of the
basic technique are described elsewhere~\cite{dcps}. 

\section{Morphological LDs with HST WFPC2}
With the advent of the high-resolution imaging capability of the Wide Field 
Planetary Camera 2 (WFPC2) on the Hubble Space Telescope (HST) the technique 
can now been
taken a step further. Instead of simply recovering the overall luminosity
function, HST's high-resolution imaging can be used to determine number-counts
and LDs according to morphology. The critical step is the
accuracy with which the morphological field number-counts are known. In
previous papers~\cite{dwg}~\cite{dwokgr}~\cite{ode} we presented detailed 
morphological number-counts from deep random field images observed by WFPC2 
(including results from the Hubble Deep Field).
Using these fields as our mean background sample we have recovered the
morphological luminosity distributions for three clusters also observed with
WFPC2. These clusters are: CL0016 (z=0.54), CL0054 (z=0.56) and CL0412 
(z=0.51), the images for which were kindly provided by the MORPHS 
collaboration~\cite{smail}. The data were 
re-analysed and morphologically classified using the same Artificial Neural
Network-based method and software as used for the field sample. This approach
to morphological galaxy classification is described in more detail by Odewahn 
{\it et al.}~\cite{ode}.

Figure 1 shows the resulting {\it mean} luminosity distributions for the
three clusters divided into three morphological categories --- E/S0, Sabc and 
Sd/Irr. The LDs have been converted to rest frame magnitudes by adopting a
mean K-correction per type and assuming a standard flat cosmology with
$H_{o}$ = 50 kms$^{-1}$Mpc$^{-1}$.

\section{Discussion}
Overlaid on Figure 1 are the local schematic morphological LDs~\cite{bst}.
The indication is for purely passive/no-evolution for early-types
and the strong evolution ($\Delta m \approx 2.5$ mags since $z = 0.55$) of 
late-type systems. This result is analogous
to that seen in the field~\cite{dwokgr} and begs the question as to whether 
the Butcher-Oemler effect~\cite{bo} seen in clusters and the Faint Blue Galaxy 
problem are one and the same ?

\section*{Acknowledgments}
SPD acknowledges DITAC for providing the funding to attend this conference
and thanks the MORPHS group for the early release of their HST WFPC2 cluster
data.

\begin{figure}[t] 
\psfig{figure=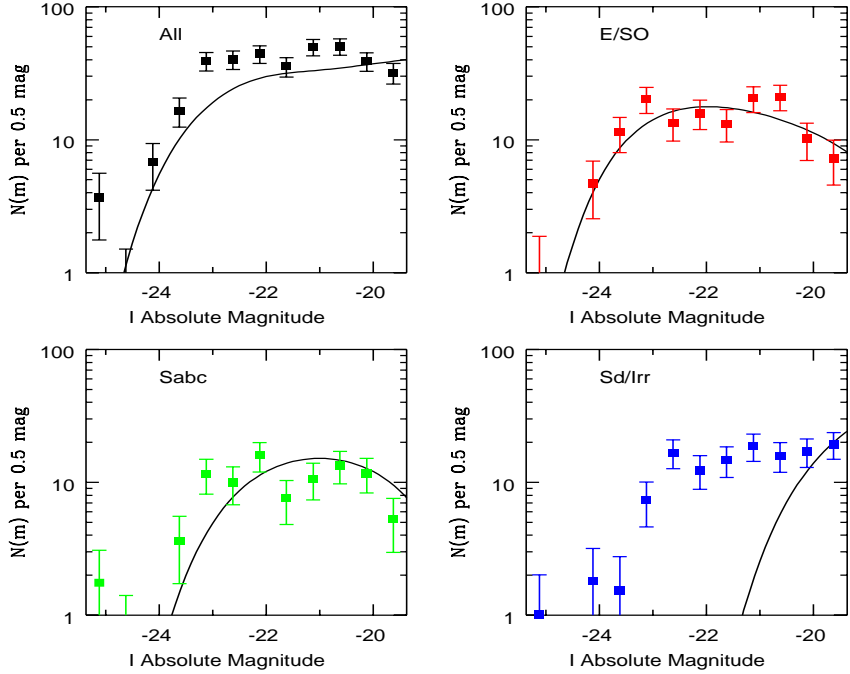,height=3.8in,width=4.75in} 
\caption{The mean morphological luminosity distributions observed in three 
medium redshift clusters. The panels represent (a) all galaxy types, (b)
E/S0s (c) Sabc's and (d) Sd/Irr's. The solid lines show the equivalent z=0
distributions based on population studies of the Virgo cluster.}
\label{fig:fig1}
\end{figure}

\end{document}